\documentclass[12pt,preprint]{aastex}
\usepackage{emulateapj5}
\def\cc{cm$^{-3}~$}
\def\ccc{cm$^{-3}$}
\def\apj{ApJ}
\def\aa{A\&A}

\def\pasp{PASP}
\def\apss{Ap\&SS}
\def\aj{AJ}
\def\mn{MNRAS}

\def\brg{Br$\gamma$\/\ }
\def\bra{Br$\alpha$\/\ }

\def\hu{H75$\alpha$\/\ }
\def\hx{H92$\alpha$\/\ }

\def\l0{{l$_0$\/\ }}

\def\s{s$^{-1}$\/\ }
\def\sss{s$^{-1}$}
\def\nlyc{N$_{\rm Lyc}$\/\ }
\def\nlycc{N$_{\rm Lyc}$}
\def\lmech{L$_{\rm mech}$\/\ }

\def\asec{$^{\prime\prime}$\/\ }
\def\asecc{$^{\prime\prime}$}

\def\msun{$M_{\odot}$\/\ }

\def\msunn{$M_{\odot}$}

\def\ha{H$\alpha$\/\ }

\def\kms{km s$^{-1}$}

\def\lsim{\raisebox{-0.3ex}{\mbox{$\stackrel{<}{_\sim} \,$}}}

\begin{document}
\title{VLA Detection of RRLs from the radio nucleus of NGC 253 : \\
       Ionization by a weak AGN, an obscured SSC or a compact SNR ?}
\author{Niruj R. Mohan}
\affil{Raman Research Institute, C.V. Raman Avenue, Sadashivanagar Post Office, Bangalore 560080, India}
\affil{and}
\affil{Joint Astronomy Program, Department of Physics, Indian Institute of Science, Bangalore 560012, India}
\email{niruj@rri.res.in}
\and
\author{K.R. Anantharamaiah$^{~1}$}
\affil{Raman Research Institute, C.V. Raman Avenue, Sadashivanagar Post Office, Bangalore 560080, India}
\altaffiltext{1}{Deceased Oct. 29, 2001.}
\and
\author{W.M. Goss}
\affil{National Radio Astronomy Observatory, PO Box O, Socorro, NM 87801, USA}
\email{mgoss@aoc.nrao.edu}
\shorttitle{RRLs from the radio nucleus of NGC 253 : AGN, SSC or SNR ?}
\shortauthors{Mohan, Anantharamaiah and Goss}

\begin{abstract}
We have imaged the \hx and \hu radio recombination line (RRL) emissions from the starburst galaxy NGC 253
with a resolution of $\sim$4 pc. The peak of the RRL emission at both frequencies coincides with the
unresolved radio nucleus. Both lines observed towards the nucleus are extremely wide, with FWHM of 
$\sim$200 km \sss. Modeling the RRL and radio continuum data for the radio nucleus shows that the 
lines arise in gas whose density is $\sim$10$^4$ \cc and mass is few thousand \msunn, which requires 
an ionizing flux of 6--20$\times$10$^{51}$ photons \sss. We consider a SNR expanding in a dense 
medium, a star cluster and also an AGN as potential ionizing sources. Based on dynamical arguments, 
we rule out an SNR as a viable ionizing source. A star cluster model was considered and the dynamics 
of the ionized gas in a stellar-wind driven structure was investigated. Such a model is consistent 
with the properties of the ionized gas only for a cluster younger than $\sim$10$^5$ years. The 
existence of such a young cluster at the nucleus seems improbable. The third model assumes the ionizing 
source to be an AGN at the nucleus. In this model, it was shown that the observed X-ray flux 
is too weak to account for the required ionizing photon flux. However, the ionization requirement
can be explained if the accretion disk is assumed to have a Big Blue Bump in its spectrum. 
Hence, we favor an AGN at the nucleus as the source responsible for ionizing the observed RRLs.
A hybrid model consisting of a inner ADAF disk and an outer thin disk is suggested, which could 
explain the radio, UV and the X-ray luminosities of the nucleus.

\end{abstract}


\section{Introduction} 

NGC 253 is a nearby (D=2.5 Mpc) spiral galaxy with the central $\sim$100 
pc hosting a vigorous starburst. The ionized gas in this region, studied by its emission
in the radio, infra-red and the optical, consists of both compact and diffuse 
components and is distributed along a highly inclined ring. 
Turner and Ho (1985)\nocite{th85} discovered a string of 
compact sources at 15 GHz, which were studied in detail by Antonucci and Ulvestad 
(1988)\nocite{au88} and Ulvestad and Antonucci (1997)\nocite{ua97}. A number 
of infra-red (IR) hotspots have also been imaged in this region (Forbes, Ward, and Depoy 
1991\nocite{forbes91}; Forbes et al. 1993\nocite{forbes93}; Pi\~na et al. 
1992\nocite{pina92}; Keto et al. 1993\nocite{keto93}; 
Sams et al. 1994\nocite{sams94}). Kalas and Wynn-Williams 
(1994)\nocite{kww94} and Sams et al (1994)\nocite{sams94} showed that most of 
these hotspots are regions of low dust extinction and are not coincident 
with the radio sources. Optical imaging by Watson et al. (1996)\nocite{watson96} 
revealed the presence of four star clusters, which were identified with individual IR knots 
(see also Forbes et al. 2000\nocite{forbes00}). The radio 
and the IR-optical sources seem to trace
different populations of objects, presumably supernova remnants and HII regions, respectively. 
Radio recombination lines (RRLs) have been observed from this galaxy in the cm (Seaquist 
and Bell 1977; Mebold et al. 1980\nocite{mebold80}; Anantharamaiah and Goss 1996\nocite{ag96}) 
and mm (Puxley et al. 1997\nocite{puxley97}) wavebands with a resolution of a few 
arcseconds (1\asecc=12 pc) or larger. Since IR and radio continuum images 
are now available with sub-arcsecond resolution, the identification of the exact 
sources of RRL emission using high resolution imaging becomes relevant.

The peak of the IR emission is offset by about 3.5\asec southwest from the 
peak of the radio continuum emission (the radio nucleus). There is no
associated radio emission towards the IR peak and weak IR emission is observed
near the radio nucleus. Keto et al. (1999)\nocite{keto99} have shown that 
the IR peak hosts a super star cluster (SSC) and they have also suggested that the radio 
nucleus is probably an AGN. \citet{th85}, based on their high-resolution VLA image of
the 15 GHz continuum emission, discovered that the radio nucleus is 
an unresolved source with high brightness temperature and suggested that 
the nucleus could harbor an AGN.
Multi-band tracers of ionized gas ([NeII]: B\"oker, Krabbe
and Storey 1998\nocite{boker98}; Keto et al. 1999\nocite{keto99}; \brg: Forbes et al. 
1993\nocite{forbes93}; optical continuum and line 
emission: Engelbracht et al. 1998\nocite{engel98}; Watson et al. 1996\nocite{watson96}; 
Forbes et al. 2000\nocite{forbes00}) show that the maximum emission at these wavebands
coincide with the position of the IR peak. It is therefore surprising that the cm-wave RRL emission 
imaged by Anantharamaiah and Goss (1996) with a resolution of 1.8\asecc$\times$1\asec 
shows that the recombination line emission peaks at the radio nucleus, with much weaker 
emission near the IR peak. In order to further investigate the characteristics of this 
emission, we have carried out sub-arcsecond observations of the RRL emission from NGC 
253 at 8.3 and 15 GHz using the VLA. These observations will help identify the compact
continuum sources from which the RRLs observed at low resolutions originate. Additionally,
since the peak of the line emission is probably coincident with the radio nucleus, the
physical properties of the ionized gas in this region, and hence the nature of the nuclear 
source can be derived by modeling the RRL emission.

\section{Observations and Results}

The 8.3 GHz \hx and the 15 GHz \hu recombination lines from NGC 253 were observed 
using the Very Large Array of the National Radio Astronomy Observatory in the A 
configuration. This observing mode yields the required parsec-scale resolution 
with the best sensitivity possible. The data analysis was done using standard 
procedures available in the software package AIPS. The datasets at 8.3 GHz, acquired 
over three periods, were concatenated for further processing. 
Of the two datasets available at 15 GHz, only one was used as the other
suffered from bandpass problems. Continuum images were made at both frequencies and are 
consistent with the images published by Ulvestad and Antonucci (1997)\nocite{ua97}. 
The channel visibilities, after hanning smoothing off-line to reduce the effects of 
Gibbs ringing, were used to construct line data-cubes. The continuum and line 
images at both frequencies were convolved to a common resolution of 
0.35\asecc$\times$0.22\asec at a P.A. of $-$10$^\circ$. The shortest baselines
in the datasets are mentioned in Table 1. The largest well-sampled angular scale
in the images are estimated to be $\sim$7.5\asec at 8.3 GHz and $\sim$3.0\asec at 
15 GHz.

The continuum emission at 8.3 GHz and 15 GHz from the central 3\asec 
region are shown in Figure 1. Also shown in the figure are 
overlays of the integrated line emission. The \hx and \hu 
spectra towards the peak of the line emission are shown in Figure 2. 
Further observational details and image parameters are listed in Table 1. 
The peak of the continuum emission corresponds to the radio nucleus and 
is associated with the source 5.79-39.0 in the compact source catalog of Ulvestad
and Antonucci (1997)\nocite{ua97}. Though the central 3\asec region contains 
multiple sources, the nucleus is separable from the surrounding
emission with our resolution of 0.3\asecc. 
The peaks of both the \hx and \hu line emission coincide with the 
unresolved radio nucleus, with much weaker emission near the IR peak. 
Extended \hx line emission is detected over a 3\asec region. 
Since the 8.3 GHz datasets were imaged with less weighting 
given to short spacing visibilities 
in order to have similar resolution as that of the 15 GHz data, 
this diffuse emission is not detectable in Figure 1. In this paper, we will only 
discuss the line emission detected against the unresolved radio nucleus.
The discussion of the extended line emission as well as the emission towards
other compact sources in the field will follow in a later paper.

\section{Modeling the RRL emission}

The properties of the ionized gas at the nucleus were modeled using the observed 
continuum and recombination lines. These observational constraints 
correspond to an area of $\sim$4 pc (one synthesized beam) and the 
values are listed in Table 1. The line emitting gas was 
assumed to be photo-ionized and the atomic level populations were derived assuming that the
gas is not in local thermodynamic equilibrium (non-LTE). The ionized gas was 
modeled as a single spherical HII region, a collection of HII regions and also
as a rectangular slab. Since 
the results are similar in all three cases, we will only discuss 
the spherical HII region model. The free parameters of the model are electron density 
(n$_{\rm e}$), diameter ($l$) of the HII region and electron temperature (T$_{\rm e}$); 
the range of parameter space is indicated in Table 2. The density was assumed 
to be uniform inside the HII region. The relative locations of both the line 
emitting thermal gas and the non-thermal continuum source along
the line of sight within the synthesized beam are unknown. Hence, the fraction of 
the non-thermal continuum radiation which is behind the thermal gas (and is responsible 
for stimulated line emission due to background radiation) is varied from zero to the 
total observed flux density in the models.
For every combination of n$_{\rm e}$, $l$ and T$_{\rm e}$, the continuum and line 
emission strengths were then calculated. Of these combinations, those models which reproduce
the observed line strengths at both frequencies and also predict continuum 
flux densities consistent with the observed values were accepted as valid solutions.
The spectral index of the unabsorbed continuum (which includes the non-thermal part 
as well the thermal part which does not contribute to the observed line emission) 
was also calculated and was constrained not to be steeper than $-$1.0.
The details of the modeling technique are discussed in detail by Anantharamaiah 
et al. (1993)\nocite{azgv93}.

The range of values for various solutions are indicated in Table 2. The results 
do not depend sensitively on the assumed value of the electron temperature within 
the range explored. Typical parameters of the line emitting gas are listed in Table 3 
for specific models. The allowed models described in Tables 1 and 2 assume that the 
unabsorbed continuum fills the synthesized beam. No solutions 
were obtained if the gas is assumed to be in LTE; the linear size of the 
region is too small and the background non-thermal radiation is too weak to 
produce a detectable RRL. Hence line enhancement due to non-LTE processes 
is essential to explain the observed line strength. Only a narrow range of 
densities (6$\times$10$^3$--1.7$\times$10$^4$ \ccc) is allowed and the ionizing 
photon rate ranges from 6--20$\times$10$^{51}$ photons \sss. A power-law multi-density
model was also considered (see Model IV of Mohan et al. 2001\nocite{rnm2001} 
for details). In this model, gas at densities which differ considerably from
the values derived in this section will not contribute appreciably to the
observerd RRL emision (see Mohan et al. 2001\nocite{rnm2001}).
Hence the constant density approximation used in the models mentioned above is 
justified. For n$_e$$<\,$6$\times$10$^3$ \ccc, the computed line emissions inside 
a $\sim$4 pc region falls short of the observed values. 
For n$_e$$>\,$1.7$\times$10$^4$ \ccc, the model predictions do not agree with 
the observed H92$\alpha$/\hu line ratio. 
The observations can be explained without invoking stimulated emission due to 
a background continuum radiation, i.e., the line emission can be explained solely
by spontaneous emission by ionized gas, which includes stimulated emission by 
its own thermal continuum.
For a given emission measure, inclusion of any externally stimulated emission 
enhances the recombination line strength compared with a case where no background
radiation is invoked.
Hence, the mass of ionized gas needed and the required ionizing photon 
flux are both less by up to a factor 2--3 for models incorporating non-thermal 
continuum radiation. If the unabsorbed continuum is assumed 
to be a distributed uniform background, then the externally stimulated 
emission can account for up to 60--70 \% of the total \hu line strength. 

If these models are computed for either the \hx or the \hu line alone, the 
range of allowed densities obtained is larger. Including only the \hx line
changes the lower limit to the allowed density to 2000 \ccc, whereas, densities
as high as 10$^5$ \cc are allowed for models which incorporate only the \hu line.
Thus, including both RRLs constrains the possible density to a much narrower 
range, assuming that all of the observed line emission at both frequencies arises
in the same region of ionized gas. 
Our models, therefore, show that the observed \hu and \hx lines arise in gas 
with a mass of a few thousand \msunn, a density of $\sim$10$^4$ \cc and require 
an ionizing photon flux of $\geq$6$\times$10$^{51}$ \sss.
The relative faintness of RRLs from near the IR peak, and the absence of a 
strong radio counterpart to the IR peak, which is a massive SSC, does 
not necessarily imply that there is less gas at the latter position; 
the gas could be of lower density instead.

\section{The ionizing source}

The RRL detection from the unresolved radio nucleus is from a sufficiently small
linear size that the interpretation is not complicated by the unknown beam filling
factors of the gas. The derived ionizing photon flux inside a $\sim$0.3\asec (4 pc) region 
can be compared with the values computed from other 
tracers of ionized gas. Using photo-ionization models, this value
corresponds to a \ha flux of 1.2$\times$10$^{-14}~$W/m$^2$. Forbes 
et al (2000)\nocite{forbes00} list the \ha fluxes of compact optical 
sources measured inside a 0.4\asec aperture. Since the strength 
of the line emission at the position of the radio nucleus is weak 
(Watson et al. 1996\nocite{watson96} and Forbes, private communication), 
we can assume that the \ha flux from this area is less than 
that of the weakest source listed in their table. Then the extinction derived 
using the \ha estimated from RRL modeling is A$_V>$14$^{\rm m}$, consistent with 
the A$_V$=24$^{\rm m}$$\pm$6 derived by Sams et al. (1994)\nocite{sams94} towards 
the radio nucleus. 
Measurements of other IR emission lines from ionized gas correspond to 
much larger apertures and since the region surrounding the radio nucleus
does contain significant amounts of ionized gas, a direct comparison is
difficult. Nevertheless, if we assume an uniform extinction of 25$^{\rm m}$
in the central region, then the extinction corrected \bra flux from
the central 6\asec (from the data of \citealt{bb84}) is a factor of 10
higher than the predictions of our model for the $\sim$0.3\asec region of the
nucleus. \citet{forbes93} measured the \brg emission inside a 2\asec aperture centered
around the radio nucleus (referred to as spot A). The \brg flux measured by them is
a factor of two higher than the value predicted by our models, this excess being
possibly due to their larger aperture. 

Since the region of interest is highly obscured, the source of the photo-ionization 
remains uncertain. We have considered three possible candidates : (1) supernova remnant 
expanding in a dense medium, (2) stellar cluster and (3) AGN. Each of these will now be 
examined in detail to investigate which of these can explain the observed properties 
of the nuclear region. The imposed constraints are the derived properties of the ionized
gas like density, size, ionizing flux and line width and also the radio, UV (derived from 
ionization) and X-ray luminosities of the nucleus. The 2--10 keV X-ray luminosity (L$_{2-10}$) of
the X-ray source coincident with the radio nucleus is 7$\times$10$^{38}$ ergs \s 
(unresolved for E $>$ 2 keV; Pietsch et al. 2001\nocite{pietsch01}, using the $XMM$--$NEWTON$).

\section{A compact supernova remnant as an ionizing source}

Chevalier and Fransson (2001\nocite{chev01}) invoked a supernova remnant (SNR)
expanding into a dense ambient medium to explain some of the compact radio sources 
in the starburst galaxy M 82 and in Arp 220 as well and we now consider such a model
to explain the nucleus of NGC 253. The high ambient density slows down the 
expansion to a few 100 \kms, confines the remnant to a smaller size, and causes it to 
quickly enter the radiative or snowplow phase. These factors could explain the spatial 
extent ($\leq$4 pc) and the line width (attributed to the expansion of the remnant) of 
the ionized gas. In the snowplow phase, the swept-up gas cools and forms a thin shell, 
and can be ionized by the X-ray photons from the hot gas inside the remnant, giving rise 
to the observed RRLs. We have calculated the dynamics and radiation of such an SNR in 
an attempt to explain the RRL observations.
In our model, the input parameters are the initial energy of the explosion, E$_{\rm o}$, 
the ambient density, n$_{\rm o}$, and the age of the compact SNR. Following the work 
of Draine and Woods (1991\nocite{dw91}), we calculate the velocity and radius of 
the shock front and also the ionization rate and the density of ionized gas in the 
shell as a function of time, E$_{\rm o}$ and n$_{\rm o}$. The time dependent value of
L$_{2-10}$ was also computed for both the Sedov phase (using the similarity solutions; 
Newman 1977\nocite{newman77}) and for the radiative phase.
The observed values (listed in Table 1 and 2) are compared with these calculated values for
the radiative phase of the SNR (when an expanding recombining gas is produced). We find that 
the models cannot simultaneously satisfy all of these constraints for any combination of 
input parameters. Hence the properties of the nucleus cannot be explained by an SNR expanding 
in a dense medium.

\section{A star cluster as an ionizing source}

In this section, we assume that the nucleus hosts a stellar cluster and derive its 
properties. The number of O stars necessary 
to produce $>\,$6$\times$10$^{51}$ photons \s is $>\,$450 and the total mass of the 
cluster is $>\,$8.5$\times$10$^4$ \msun (computed assuming a Salpeter IMF with a mass 
range between 1--80 \msun and using the tables in Vacca, Garmany and Shull 
1996\nocite{vgs96}). Since the region modeled is $\sim$4 pc in extent, the resultant
total stellar mass and the lower limit to the stellar surface density imply that the 
ionizing source must be a super star cluster (SSC; Meurer et al 
1995\nocite{meurer95}). The RRLs are assumed to arise in the HII region around this SSC
(similar to `supernebulae' discovered in centers of other galaxies: Kobulnicky and 
Johnson 1999\nocite{kj99}; Neff and Ulvestad 2000\nocite{nu00}; Tarchi et al. 
2000\nocite{tarchi00}; Turner, Beck and Ho 2000\nocite{tbh00}; Mohan et al. 
2001\nocite{rnm2001}). The detection of wide lines with FWHM of $\sim$200 km \s 
from within a $\sim$4 pc region is a constraint on the dynamics of the nebula, 
which we investigate below. It should be noted that the dynamical age of the gas 
($\sim$radius/velocity) is $\sim$10$^4$ years and the sound crossing time is 
$\sim$10$^5$ years. 

\subsection{Gas dynamics of the `supernebula'}

Though a HII region ionized by a central cluster will 
expand outwards, its expansion velocity can never exceed the speed of sound in 
the ionized medium (Spitzer 1968\nocite{spit68}) and hence cannot account for 
the wide lines observed. Also champagne flows in HII regions cannot produce 
velocities larger than $\sim$50 km \s (Yorke, Tenorio-Tagle and Bodenheimer 
1984\nocite{yorke84}). Though supersonic velocities are known in extragalactic 
HII regions, their line widths are only $\sim$30--100 km \s and the HII regions 
are also much larger in size (for eg., Mu\~noz-Tu\~n\'on, Tenorio-Tagle and 
Caste\~nada 1996\nocite{mtttc}). The only plausible way to explain the observed 
velocity width is by invoking a stellar-wind driven HII region expanding at $\sim$100 
km \sss.  The expansion velocity of a wind-driven shell scales as (\lmech)$^{1/5}$, 
where \lmech is the mechanical luminosity of the wind and hence a expansion velocity of
100 km \s for a cluster whose \lmech is $\sim$10$^3$ times that of an O star is easily
obtainable. The dynamical age of a wind-driven shell is 
$\sim$0.55$\times$R$_{shell}$/V$_{shell}$ (McCray 1983\nocite{cray83}), i.e., 
$\sim$10$^4$ years. 

The model considered here is that of a uniform density nebula whose dynamics is 
determined by the ionizing photons and the stellar wind of a central SSC. In the
wind-driven phase of expansion, the nebula quickly enters the snowplow phase 
($\leq$10$^3$ years), wherein it can be either in the energy-conserving or in the
momentum-conserving phase. The swept-up matter forms a thin shell which is assumed
to trap all the ionizing photons. The input parameters are the star formation history
of the cluster, the constant ambient density (n$_{\rm o}$) and the age of the cluster 
(assumed to be the same as the nebula) and the range of parameter space explored is 
summarized in Table 4. The time dependent values of \nlyc
and \lmech were derived from simulations using 
the code Starburst 99 \footnote{URL: http://www.stsci.edu/science/starburst99/} 
(Leitherer et al. 1999\nocite{sb99}). This simulation was carried out for solar 
metallicity and a Salpeter initial mass function with an upper mass cut-off of 100 
\msunn. These values were obtained both for a continuous and an instantaneous star 
formation history for a particular value of star formation rate (SFR) and subsequently 
scaled linearly for other values of SFR. For a given combination of input parameters,
the radius and velocity of the shock front (almost the same as that of the ionized gas),
mass of ionized gas, ionization rate and the gas density were computed.
These calculations were carried out using the expressions in Shull (1980)\nocite{shull80} 
for the appropriate phase of the nebula. The derived values were compared with the 
values derived from the RRL modeling, listed in Table 4. Acceptable solutions were 
identified which were consistent with the imposed constraints. Table 4 summarizes the 
allowed parameters. 

Gas dynamics narrowly constrains the properties of the possible supernebula.
The derived solutions correspond to wind-driven shells in the energy-conserving phase.
The allowed age of the nebula (and hence the cluster) is $\sim$2.5$\times$10$^4$ years, corresponding 
to a diameter of 7--8 pc; no solutions for smaller sizes were found. These values are a 
strong constraint on the hypothesized SSC. However, the reliability of these values 
depends on the accuracy of the Starburst 99 code for young clusters, as discussed 
in section 8.1.

\section{A weak AGN as an ionizing source}

\citet{th85} discovered that the nucleus is unresolved in the radio continuum, 
and has a high brightness temperature and hence suggested a possible AGN at
the nucleus. \citet{ua97} also arrived at the same conclusion based on their 
higher-resolution images. The derived size is $<$ 0.05\asec and the lower limit
to the brightness temperature is between 22\,000-90\,000 K. If an AGN does exist 
at the nucleus, then
its UV continuum can be invoked to ionize the RRL emitting
gas. The expected UV luminosity can be estimated using the observed radio continuum 
and X-ray luminosities. Based on the upper limit to the size of the ionized gas 
and the observed line width, the dynamical mass at the center ($\sim$$\sigma^2r$/G) 
is constrained to be less than 3$\times$10$^6$ \msunn. 

Powerful AGNs are powered by a geometrically thin standard accretion disk whose
blackbody emission peaks in the UV, called the Big Blue Bump, or BBB (Koratkar 
and Blaes 1999\nocite{kb99}), which can provide the required ionization. Falcke 
et al. (1995\nocite{falcke95}) showed that this BBB luminosity is correlated with 
the core radio emission ($\nu$L$_\nu$ at 2 cm) for a sample of quasars. Extrapolating 
their relation for radio-weak quasars for the 2 cm luminosity of NGC 253
which is about hundred times less than their sample, the BBB luminosity for the nucleus 
is estimated to be $\sim$10$^{42}$ ergs \sss. This UV luminosity corresponds 
to an ionizing photon flux of $\sim$2$\times$10$^{52}$ \sss, consistent with the value derived
from the observed RRLs. Based on the derived UV luminosity, the value of L$_{2-10}$
was calculated for a range of input parameters (Frank, King and Raine 1985\nocite{fkr85}).
It was found that the thin disk model under-predicts L$_{2-10}$, and is consistent with
observations only for super-eddington accretion rates.

However, given the low radiative efficiency of the nucleus (the accretion rate is 
estimated to be less than 0.01 times the eddington rate), the AGN could be powered
by an ADAF disk (Narayan, Mahadevan and Quataert 1998\nocite{nmq98}). ADAF models 
predict the absence of a BBB. The 2 cm radio luminosity 
of the nucleus of NGC 253 implies that the AGN is of low luminosity (LLAGN; Nagar 
et al. 2000\nocite{nagar00}) and these LLAGNs have been shown not to exhibit a BBB 
in their UV emission (Ho 1999\nocite{ho99}). Hence we shall now try to explain the 
ionization from the X-ray 
luminosity instead. Pietsch et al. (2001\nocite{pietsch01}) modeled the emission from 
the nuclear X-ray source as bremsstrahlung radiation from a three-temperature plasma. 
For simplicity, we shall assume that the UV up to X-ray emission is bremsstrahlung 
radiation from plasma at a single temperature, T$_{\rm x}$. The UV portion of the emission 
is constrained to produce an ionizing photon flux greater than that needed to explain the 
RRLs and the 2--10 keV portion is normalized to the observed value. The required ionization is
found to be consistent with the observed X-ray flux only for T$_{\rm x}$$<$0.2 keV. Since it is 
unlikely that most of the 2--10 keV emission arises in the exponentially decaying part of the
thermal emission at 0.2 keV, this result is interpreted to imply that the X-ray emission is too weak
to explain the required ionization. 

\section{Discussion}

\subsection{The star cluster model}

Given the uncertainty of stellar tracks, \nlyc and \lmech predicted by the Starburst 99 code are 
not accurate to more than a factor of a few for t$<$5$\times$10$^5$ years (Leitherer, private 
communication). The radius of the shock front is
R$_{\rm s}$$\sim$($\int{{\rm L_{mech}} dt}$)$^{1/5}$ in the energy-conserving phase. Since 
\lmech is a monotonically increasing function of time (up to 5$\times$10$^6$ years for a 
continuous SFR), the dynamics of the nebula at t$>$5$\times$10$^5$ years is insensitive to 
the uncertainties in the Starburst 99 model at earlier times. Hence we can conclude that a 
nebula older than this age will not be able to explain the observations. The only consistent 
nebular age estimate derived in section 6.1 is $\sim$2.5$\times$10$^4$ years. The inputs from 
the Starburst 99 code are not accurate for such short timescales. Nevertheless it can be shown 
that since the ratio between \lmech and \nlyc is constant with time, if just the functional 
dependence of \lmech with time for t$<$5$\times$10$^5$ years is reliable, then the model 
results are relatively independent of the uncertainties in the Starburst 99 codes. However, 
given the short ages derived, it can be expected that the star formation at this timescale, 
and hence the value of \lmech and \nlycc, is stochastic, in which case, the derived age is 
inconsistent with the assumptions in the model. Even for a stochastic star formation process, 
the short dynamical timescale of 10$^4$ years needs to be explained. The dynamical model 
implicitly assumes that the star cluster is spatially within the wind-blown structure, which
need not be true for a $\leq$4 pc region. The model also assumes that the stellar wind of 
the individual stars add coherently to produce a total `cluster-wind'. This assumption will
be violated for a young cluster due to stellar-wind collisions and YSO outflows. These
factors may help confine the gas for a longer time and also provide a natural way of increasing 
the line width by stirring up the intra-cluster medium. Hence if a star cluster is invoked as
an ionizing source, it must be fairly young, though the properties of such a young cluster
cannot be easily predicted.

\subsection{The AGN model}

We have shown in section 7 that if the proposed AGN hosts a standard thin accretion disk 
with a BBB, then this disk produces enough UV radiation to ionize the RRL emitting gas but
its X-ray emission falls short of the observed value. On the other hand, if the AGN has an 
ADAF disk instead, then again, the observed X-ray emission is insufficient to ionize
the RRL emitting gas. ADAF models also predict the X-ray emission for a given radio flux and
mass of the black hole (Yi and Boughn 1998\nocite{yb98}) and the estimated X-ray luminosity
is a hundred times less than that observed (and is ten times less than the ratio observed in
other LLAGNs; Ulvestad and Ho 2001\nocite{uh01}). A possible way of explaining the observed
radio and X-ray fluxes along with the required UV flux is the model of Quataert et al. 
(1999\nocite{qua99}) who explain the spectrum of two LLAGNs which have an X-ray to 
UV ratio which is too large for
a thin disk and too small for an ADAF disk. Following their work, we hyphothesize that the
possible AGN in the nucleus of NGC 253 has an ADAF disk in the interior, giving way to a 
standard thin disk beyond a certain radius. In this picture, the radio and the UV emissions 
arise in the outer disk (which would have a BBB in its spectrum) and the X-ray emission arises
from the inner ADAF disk. 

\subsection{The ionizing source}

It is clear from section 8.1 that if the ionizing source is a star cluster, its age must 
be atleast younger than $\sim$10$^5$ years. The AGN model, discussed in section 8.2, does
manage to explain the observations but the constraints on this model are relatively less in
number. The probability of detecting a star cluster of age $\lsim$10$^5$ years
at the radio nucleus would be small. Since there is additional evidence supporting the existence
of an AGN at the nucleus, we favor an AGN as the possible ionizing 
source. The existence of an AGN at the nucleus needs to be confirmed through high
resolution radio continuum or X-ray imaging. The surrounding region has a transitional 
HII/weak-[O I] LINER spectrum (Engelbracht et al. 1998\nocite{engel98}) and hence
disentangling the optical signature of an AGN from this LINER emission will be
difficult. Detailed multi-wavelength modeling of the nuclear emission also needs to 
be carried out in the framework of the AGN model in order to further constrain the
properties of this object.

\section{Conclusions}

We have imaged the RRL emission from the starburst galaxy NGC 253 at 8.3
GHz and 15 GHz with a spatial resolution of $\sim$0.3\asec ($\sim$4 pc)
using the VLA. The line emission is maximum at the radio nucleus at both frequencies and
is much weaker near the position of the IR peak, which is known to host an SSC. The line widths
of both RRLs are large: $\sim$200 km \sss. The continuum and
line emission were modeled in terms of a uniform density photo-ionized gas.
The observed RRLs can be explained as arising from a 2--4 pc sized
region of gas of mass few thousand \msunn, at a density of 10$^4$ \ccc. The
ionizing flux required is 6--20$\times$10$^{51}$ photons \sss. This gas
can, in principle, be ionized by a compact SNR, a star cluster or by an AGN. The dust
extinction against the nucleus is very high and hence direct detection of the
ionizing source in the optical-IR is not feasible. Detailed dynamical modeling
shows that a compact SNR cannot explain the observed properties of the ionized gas.
The star cluster model was investigated in terms of a stellar wind-blown structure.
Such a model can account for all of the observed properties only for a cluster of
age $\sim$2.5$\times$10$^4$ years. Though this age estimate is shown not to be 
consistent with the dynamical model considered, a relatively young age ($\lsim$10$^5$ 
years) for any such cluster is unavoidable. If an AGN is assumed to be the ionizing
source, then the observed X-ray flux cannot explain the required ionizing photon rate. 
However, a simple thin accretion disk model can account for the ionization based on the
observed radio flux density. A composite model involving an inner thin disk and an outer
ADAF disk is suggested, in order to simultaneously explain the radio, UV and the X-ray
observations.

The detection of such a young star cluster at the nucleus of the galaxy is 
improbable. Since there is additional evidence supporting the existence of an AGN at the 
nucleus from radio continuum data, we favor the AGN model as an ionizing source. 
If confirmed, our observations could be the first 
detection of RRLs from an AGN outside our galaxy. The central $\sim$36 pc region of NGC 253 
is host to a number of compact thermal and non-thermal sources and diffuse gas and is also 
coincident with the base of the galactic superwind and possibly hosts a central AGN as well. Hence 
this galaxy would be an ideal laboratory to study the dynamics and the interaction of all
these components, much like the galactic center.

\acknowledgments The National Radio Astronomy Observatory is a facility of the
National Science Foundation operated under cooperative agreement by Associated
Universities, Inc. This research has made use of the Starburst 99 code, 
made available to the public by the authors and we thank them for sharing it 
with the community. This research has made use of NASA's Astrophysics Data System 
Abstract Service. We thank Dipankar Bhattacharya, K. S. Dwarakanath, Kelsey Johnson, 
Biman Nath, Prasad Subramanian and Jim Ulvestad for detailed discussions and 
helpful comments. We also wish to thank the referee for many insightful comments
which helped in improving the clarity of the paper.

{}

\begin{deluxetable}{lcc}
\tablecolumns{3}
\tablewidth{0pc}
\tablenum{1}
\tablecaption{\bf VLA observational log and image parameters}
\tablehead{
\colhead{Parameter}& \colhead{8.3 GHz data}  & \colhead{15 GHz data} }
\startdata
Date of observation             & 9/7/99, 12/7/99  & 26/6/99 \\
                                & \& 9/10/99 & \\
$\nu_{\rm rest}$ of RRL (GHz)   & 8309.4 (H92$\alpha$)  &  15281.5 (H75$\alpha$) \\
Bandwidth (MHz), channels/IF & 24.2, 31 & 46.9, 15\\
Spectral resolution\tablenotemark{a}\,\,\,\,\,(km \sss)   & 56.4 & 122.6 \\
Shortest baseline (k$\lambda$)  & 7 & 12 \\
Beam (natural weights)          & 0.5\asecc$\times$0.28\asec & 0.31\asecc$\times$0.14\asec \\
Phase calibrator                & 0116-219 & 0118-272 \\
Bandpass calibrator             & 2251+158 & 2251+158 \\
\hline
Peak continuum flux density\tablenotemark{b}\,\,\,\,\,(mJy) & 40 & 37  \\     
Peak line flux density (mJy)   & 0.92$\pm$0.02  & 1.72$\pm$0.02 \\
Noise in the continuum image (1$\sigma$, mJy) & 0.04 & 0.13 \\
Line strength\tablenotemark{c}\,\,\,\,\,($\times$10$^{-23}$ W/m$^2$)  & 6$\pm$1  & 23$\pm$2  \\
FWHM of line\tablenotemark{d}\,\,\,\, (km \sss)          & 225 $\pm$ 15 & 197 $\pm$ 26 \\ 
Noise per channel (mJy)         & 0.14 & 0.35 \\
\enddata
\tablenotetext{a}{The spectral resolution after off-line hanning smoothing.}
\tablenotetext{b}{Evaluated by simultaneously fitting a gaussian, a zero level and 
a slope.}
\tablenotetext{c}{Corresponding to the spectrum against the unresolved radio nucleus.}
\tablenotetext{d}{The FWHM after deconvolving the effects of hanning smoothing
and finite spectral resolution.}
\end{deluxetable}

\begin{deluxetable}{lc}
\tablecolumns{2}
\tablewidth{0pc}
\tablenum{2}
\tablecaption{\bf RRL model details (spherical geometry)}
\tablehead{
\colhead{Parameter}& \colhead{Range} \\
\hline
\multicolumn{2}{c}{\em Explored input parameter space}}
\startdata
Electron temperature (T$_{\rm e}$, K)   & 2500--12500 \\
Local electron density (n$_e$, \ccc)    & 10$^{-2}$--10$^6$ \\
Diameter ($l$, pc)                      & 0.01--5.0 \\
\cutinhead{\em Model solutions for T$_{\rm e}$=5000--12500 K}
n$_e$ (\ccc)             & 6000--17000 \\
$l$   (pc)               & 2--5.0 \\
\nlyc (\sss)             & (6--20)$\times$10$^{51}$\\
M$_{\rm HII}$ (\msunn)  & 1000--7000\\
Stimulated emission (at 15 GHz, \%)     & 10--60 \\
Thermal continuum fraction (15 GHz) & 0.15--0.7 \\
\enddata
\end{deluxetable}

\begin{deluxetable}{lccccccc}
\tablecolumns{8}
\tablewidth{0pc}
\tablenum{3}
\tablecaption{{\bf Typical model results :} {\rm for a spherical 
geometry with T$_{\rm e}$=7500 K.}}
\tablehead{
\colhead{Parameter}& \multicolumn{3}{c}{n$_{\rm e}$=7000 \ccc} &  & \multicolumn{3}{c}{n$_{\rm e}$=17000 \ccc} \\
\cline{2-4}
\cline{6-8}
 &  \colhead{A\tablenotemark{a}} & \colhead{B} & \colhead{C} & & \colhead{A} & \colhead{B} & \colhead{C}} 
\startdata
Diameter $l$ (pc)      & 3.1 & 4.0 & 3.5 &~~~& 2.1 & 2.3 & 2.1 \\
N$_{\rm Lyc}$ (\,$\times$10$^{52}$ \sss)   & 0.8 & 1.6 & 1.0 &~~~& 1.4 & 1.9 & 1.4 \\
S$_{\rm th}$/S$_{\rm total}$ at 15 GHz  & 0.3 & 0.6 & 0.4 &~~~& 0.4 & 0.5 & 0.4 \\
Stimulated emission\tablenotemark{b}~~(\%) & 55 & 0 & ---  &~~~& 22 & 0 & --- \\
Spectral index\tablenotemark{c}~~(S$_{\nu}$\,$\propto$\,$\nu^{-\alpha}$) & 0.45 & 0.5 & 0.45 &~~~& 0.7 & 0.75 & 0.7 \\
M$_{\rm HII}$ (\msunn) & 2800 & 6000 & 3700 &~~~& 2100 & 2800 & 2200 \\
\enddata
\tablenotetext{a}{A, B, and C refer to models where the unabsorbed
continuum radiation is assumed to be behind, in front of, and mixed with, 
the ionized gas.}
\tablenotetext{b}{The fraction of \hu line emission due to stimulated emission
by the background radiation.}
\tablenotetext{c}{Average spectral index of the unabsorbed continuum radiation, excluding the
thermal contribution of the ionized gas.}
\end{deluxetable}

\begin{deluxetable}{lcc}
\tablecolumns{3}
\tablewidth{0pc}
\tablenum{4}
\tablecaption{\bf Dynamical modeling details}
\tablehead{
\colhead{Parameter}& \colhead{Explored Range} & \colhead{Solutions obtained}\\
\hline
\multicolumn{3}{c}{\em Parameter space for model inputs}}
\startdata
Ambient density (n$_{\rm o}$, \ccc)              & 10$^{-3}$--10$^6$  & 200--400 \\
Age of cluster (yrs)                             & 10$^4$--10$^8$  & (2--3.5)$\times$10$^4$ \\
SFR (\msunn/yr)                                  & 0.1--100.0  & 2.5--6 \\
\cutinhead{\em Parameter space for observational constraints}
Radius of shell (R$_{\rm s}$, pc)                & 0.2--4.0 & 3.5--4.0 \\
Velocity of shell (V$_{\rm s}$, km \sss)         & 80--120 &  85--115 \\
Mass of ionized gas (\msunn)                     & 1250--5000  & 1250--2400 \\
\nlyc (photons \sss)                             & 6--24$\times$10$^{51}$ & 6--10$\times$10$^{51}$ \\
Density of ionized gas (n$_e$, \ccc)             & (4--25)$\times$10$^3$ & (18--25)$\times$10$^3$ \\
\enddata
\end{deluxetable}

\begin{figure}
\figurenum{1}
\epsscale{0.9}
\plottwo{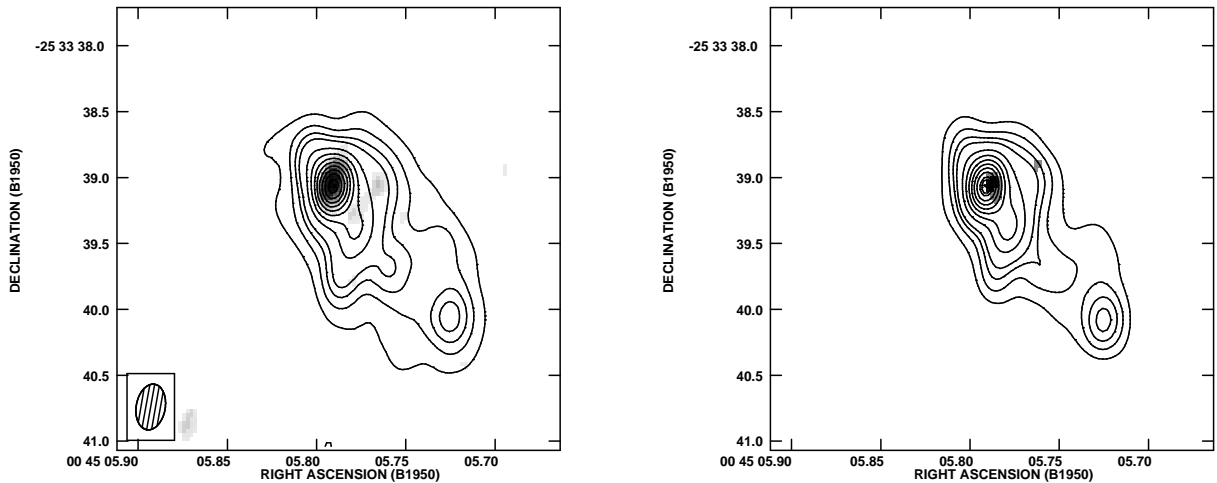}{f1b.ps}
\caption{The radio continuum emission at 8.3 GHz ({\em left}) and 15 GHz ({\em right}) 
of the central 36 pc of NGC 253 are plotted as contours. The contour levels in mJy are 
4, 6, 8, 10 and then higher in steps of 5 mJy. The line emission integrated over the 
velocity range 100--400 km \s for the \hx and \hu RRLs are overlaid in greyscale.
The beam size for both the images is 0.35\asecc$\times$0.22\asec at a P.A. of 
$-$10$^\circ$ and is shown at the bottom left corner of the left-hand side image.}
\end{figure}

\begin{figure}
\figurenum{2}
\epsscale{0.9}
\plottwo{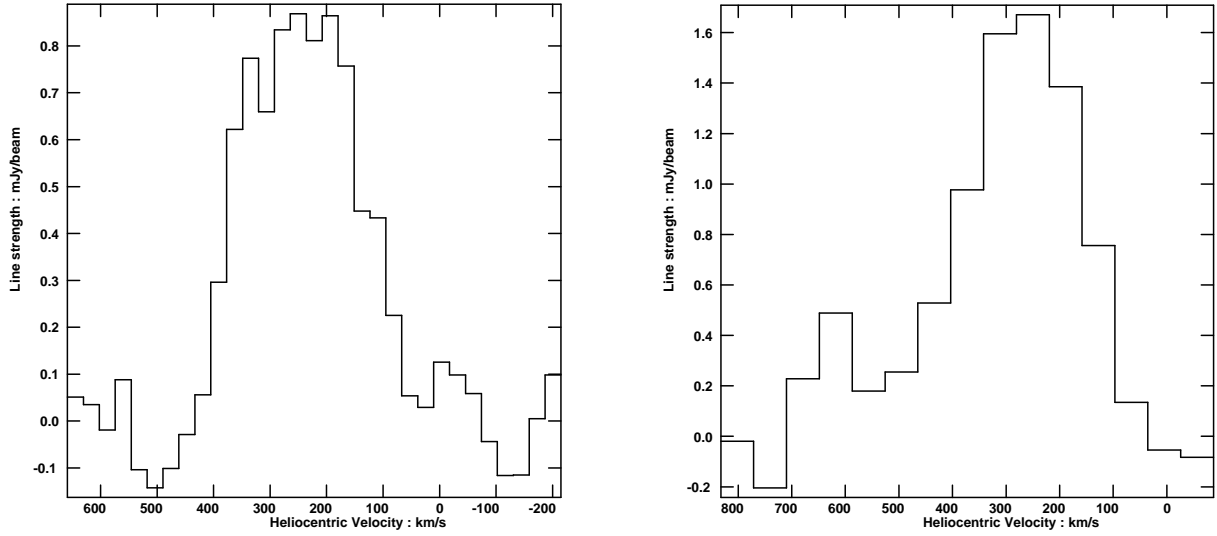}{f2b.ps}
\caption{Hanning smoothed spectra of \hx RRL at 8.3 GHz ({\em left}) and \hu RRL at 15 GHz ({\em right}) 
against the radio nucleus, for a resolution element of 0.35\asecc$\times$0.22\asecc. The velocity 
resolution is 56.4 km \s and 122.6 km \s for the \hx and \hu spectra respectively.}
\end{figure}

\end{document}